\documentclass[preprint,prl,aps,amsmath,amssymb,color,superscriptaddress]{revtex4}
\usepackage{stmaryrd}
\usepackage{amsmath}
\usepackage{amssymb}
\usepackage{graphicx}
\usepackage{dcolumn}
\usepackage{bm}
\usepackage{multirow}
\usepackage{hyperref}
\usepackage{color}
\begin{document}

\begin{titlepage}
\title{Understanding the origin of bandgap problem in transition and post-transition metal oxides}
\author{Hengxin Tan}
\affiliation{State Key Laboratory of Low-Dimensional Quantum Physics and Collaborative Innovation Center of Quantum Matter, Department of Physics, Tsinghua University, Beijing 100084, China}
\author{Haitao Liu}
\affiliation{Institute of Applied Physics and Computational Mathematics, PO Box 8009, Beijing 100088, China}
\author{Yuanchang Li}
\email{yuancli@bit.edu.cn}
\affiliation{Key Lab of advanced optoelectronic quantum architecture and measurement (MOE), and Advanced Research Institute of Multidisciplinary Science, Beijing Institute of Technology, Beijing 100081, China}
\author{Wenhui Duan}
\affiliation{State Key Laboratory of Low-Dimensional Quantum Physics and Collaborative Innovation Center of Quantum Matter, Department of Physics, Tsinghua University, Beijing 100084, China}
\affiliation{Institute for Advanced Study, Tsinghua University, Beijing 100084, China}
\author{Shengbai Zhang}
\affiliation{Department of Physics, Applied Physics and Astronomy, Rensselaer Polytechnic Institute, Troy, NY, 12180, USA}
\date{\today}

\begin{abstract}
  Improving electronic structure calculations for practical and technologically-important materials has been a never-ending pursue. This is especially true for transition and post-transition metal oxides for which the current first-principles approaches still suffer various drawbacks. Here we present a hierarchical-hybrid functional approach built on the use of pseudopotentials. The key is to introduce a discontinuity in the exchange functional between core and valence electrons. It allows for treating the localization errors of $sp$ and $d$ electrons differently, which have been known to be an important source of error for the band structure. Using ZnO as a prototype, we show the approach is successful in simultaneously reproducing the band gap and $d$-band position. Remarkably, the same approach, without having to change the hybrid mixing parameters from those of Zn, works reasonably well for other binary $3d$ transition and post-transition metal oxides across board. Our findings point to a new direction of systematically improving the exchange functional in first-principles calculations.
\end{abstract}

\maketitle
\draft
\vspace{2mm}
\end{titlepage}

Transition and post-transition metal oxides are among the most popular class of inorganic solids as they show many interesting physical properties including, among others, metal-insulator transition, magnetism, ferroelectricity, colossal magnetoresistance, charge order, and high temperature superconductivity \cite{Rao,MIT,Science288p462}. They are also technologically important for numerous applications such as catalysis, gas sensors, and electro-/photo-/thermochromic devices \cite{Nature453p80,Kung,AM24p5408}. Understanding the vastly-diverse behaviors of these metal oxides requires an adequate description of their underlying electronic structure.

First-principles methods are routinely used to study electronic structure of solids from which one can obtain mechanical, electrical, and optical properties. Density functional theory (DFT) \cite{Kohn,Jones} is one of the most employed such approaches. Although DFT has achieved great successes in the past, it runs into difficulties for transition and post-transition metal oxides due to the challenge in dealing with the localized $d$ or $f$ electrons \cite{PRL65p1148,Cohen}. Self-interaction has been blamed for the errors as a result of an over-delocalization of the electrons. This leads to a too-small band gap ($E_g$) and a too-high $d$-band energy ($E_d$) relative to the valence band maximum (VBM). Hartree-Fock (HF) approach, on the other hand, overly localizes the electrons, giving rise to errors in the opposite direction of the DFT, namely, it overestimates $E_g$ while produces a too low $E_d$ with respect to the VBM \cite{Sanchez}. As a logical choice, one may mix the DFT with HF, i.e., in a hybrid approach, to improve the numerical accuracy.
Such hybrid approaches are within generalized Kohn-Sham scheme whose single-particle eigenvalue gaps already incorporate part of the discontinuity of the functional derivative of the exchange-correlation energy \cite{PRB53p3764,PNAS114p2801,RMP80p3}. Therefore, it may also hold the potential to reproduce the experiment $E_g$.
Although working well for the $sp$-electron systems, the hybrid functional, e.g., the HSE \cite{JCP118p8207,JCP125p224106}, runs into difficulties for the transition and post-transition metal oxides \cite{PRB88p041107,JCC38p781}.
What is the fundamental difference between $sp$ and $sp$-$d$ mixed systems from the viewpoint of electronic structure calculations? Can one obtain satisfactory band structure for $sp$-$d$ mixed systems within the same hybrid scheme as that for $sp$ only systems? These questions are critical to the prevalently used hybrid functional methods, as well as the understanding of modern electronic structure theory.

In this paper, we first consider ZnO\ ---\ a notoriously bad player among semiconductors with awfully-large errors for both $E_g$ and $E_d$ up to several electron volts (eV). We show that the conventional all-electron HSE with a single mixing parameter $\alpha$ cannot simultaneously reproduce the experimental $E_g$ and $E_d$. The reason is because the orbital-dependent localization errors are spatially inhomogeneous making the homogeneous hybrid scheme ineffective.
We introduce a {\it hybrid functional pseudopotential} (PP) \cite{PRB97p085130} based hierarchical-hybrid functional (HHF) approach for the electronic structure of transition and post-transition metal oxides.
By using different hybrid functionals, i.e., different HF mixing parameters $\alpha_c$ and $\alpha_v$, to treat the core and valence electrons of the metal elements, the approach can account for the spatial inhomogeneity of the localization errors. At ($\alpha_c$, $\alpha_v$) = (0.75, 0.25), it simultaneously reproduces the experimental $E_g$ and $E_d$ for ZnO. More intriguing is the fact that the method, with the same ($\alpha_c$, $\alpha_v$) = (0.75, 0.25) works for 11 other binary 3$d$ transition and post-transition metal oxides as well, especially for Cr$_2$O$_3$, MnO, Fe$_2$O$_3$ and CuO, which are also notoriously-bad examples for HSE.

A hybrid functional is obtained by mixing PBE and HF as follows
\begin{equation}\label{(1)}
   E_{xc}^\textrm{hybrid} = \alpha E_{x}^\textrm{HF} + ( 1 - \alpha ) E_{x}^\textrm{PBE} + E_{c}^\textrm{PBE},
\end{equation}
where the mixing parameter $\alpha$ specifies the amount of HF exchange $E_x^\textrm{HF}$ to replace the PBE functional \cite{PRL77p3865}. If $\alpha$ = 0, Eq. (1) is reduced to the PBE functional; if $\alpha$ = 1, on the other hand, it becomes 100\% HF, while the correlation functional remains to be 100\% PBE. When $\alpha$ = 0.25, it is known as the PBE0 functional \cite{JCP105p9982,JCP110p6185}. The widely-used HSE functional is obtained by screening off the long-range tail of HF exchange in PBE0.

Our calculations were performed using the Quantum ESPRESSO \cite{QE} with a kinetic-energy cutoff of 60 Ry.
All metal PPs were constructed by the OPIUM \cite{OPIUM} code (See Supplemental Material \cite{SM} as well as Ref. \onlinecite{tanPCCP} for details). As our focus here was on electronic structure, we used experimental lattice parameters for the oxides. For comparison, we also performed all-electron calculations using the FHI-aims code \cite{FHI-aims}. Magnetic structures used in calculations can be found in Supplemental Material \cite{SM}.

\begin{figure}[tbp]
\includegraphics[width=0.75\columnwidth]{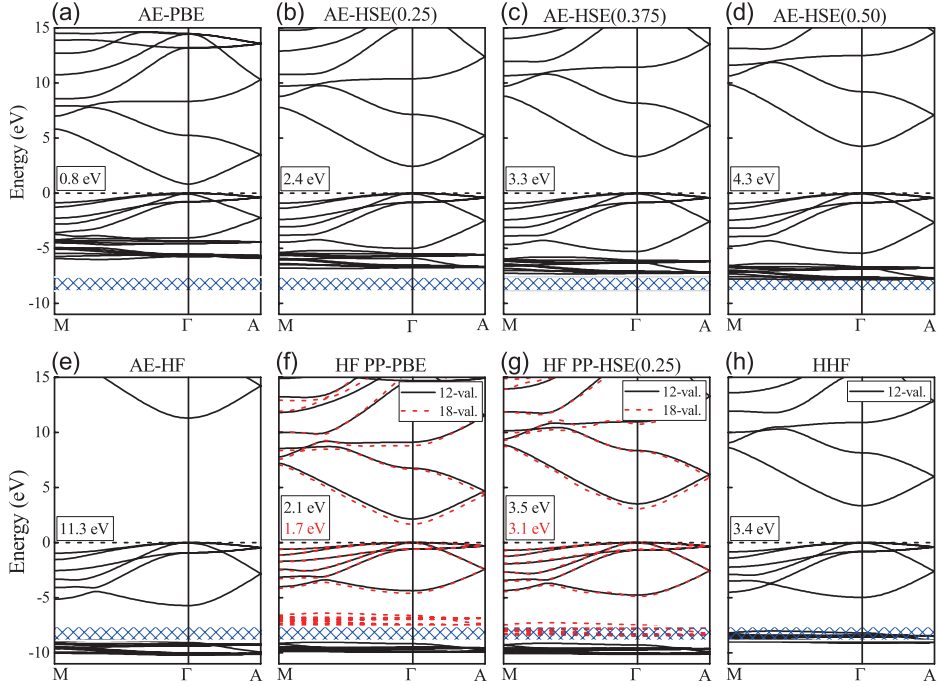}
\caption{Band structures of ZnO. Panels (a) to (e) are the all-electron (AE) results with different exchange functional forms: (a) PBE, (b)$-$(d) HSE with different $\alpha$ values (0.25, 0.375, 0.50), and (e) HF. Panels (f) and (g) are the HF PP results where (f) is PBE and (g) is 25\% HSE for the valence electrons, respectively. In both panels (f) and (g), two different Zn HF PPs are considered with 12- and 18-valence electrons (12-val. and 18-val.), respectively. (h) Our HHF method with ($\alpha$$_c$, $\alpha$$_v$) = (0.75, 0.25). More details can be found in text. In the plots, black dashed lines denote the position of the VBM; the framed numbers denote the $E_g$, while blue grids denote experimental $E_d$. For clarity, the calculated $E_d$ is the average $d$-band position below VBM.}
\end{figure}

Figure 1 shows the band structures of ZnO, which is used as the benchmark system in the exploration of various functional forms. Following Ref. \cite{WeiSH}, we consider $E_g$ and $E_d$ as the two single-most important physical parameters for electronic structure. Experiment showed that $E_g$ = 3.4 eV \cite{Madelung}, while $E_d$ is located in the range of 7.5 $\sim$ 8.8 eV below the VBM \cite{PRL27p97,JPCM17p1271,SSC54p701,JAP76p5495,PRB5p2296,PRB4p451,PRB9p600}. More specifically, Figs. 1(a)$-$(e) show the all-electron results: (a) PBE, (b) HSE with $\alpha=0.25$, (c) HSE with $\alpha$ = 0.375, (d) HSE with $\alpha$ = 0.5, and (e) HF. One sees that, as $\alpha$ increases from 0 (i.e. PBE) to 0.5, $E_g$ increases quickly while $E_d$ decreases accordingly but to a much lesser degree. At the default $\alpha$ = 0.25 [Fig. 1(b)], $E_g$ of 2.4 eV and $E_d$ of $-$6.0 eV deviate from experiment by more than 1 eV. Increasing $\alpha$ to 0.375 can reproduce the experimental $E_g$ but $E_d$ is still considerably away from experiment [Fig. 1(c)]. At $\alpha$ = 0.5, the $d$ bands approach the upper bound of the experimental value [Fig. 1(d)]. However, the corresponding $E_g$ of 4.3 eV is too large when compared to experiment. At all-electron HF in Fig. 1(e), the $d$ bands with an $E_d$ = $-$9.7 eV are too deep and the $E_g$ of 11.3 eV is also too large. These all-electron results reveal the inability of the single $\alpha$ hybrid approach, and also indicate that a much larger amount of HF is required to correct $E_d$ than what is desired to correct $E_g$.

In the current implementation of the hybrid functional approaches, a same amount, e.g., 25\% of HF \cite{JCP105p9982} has been used in HSE throughout. This amounts to a mapping of the orbital-dependent localization error of the DFT onto a spatially homogeneous reference system. Such a single-parameter treatment seems to be fine when the inhomogeneity of the errors is not significant, as in the $sp$ systems. However, the $d$-electron states inherently possess a larger DFT localization error than the $sp$-electron states, and it is physically as important as the $sp$-electron for transition and post-transition metal oxides \cite{WeiSH}.
In such a case, it is clear that a single-parameter treatment is insufficient.
At least two parameters are needed to mix the exact exchange, respectively, for the $sp$- and $d$-electrons. Our simple argument here not only applies to ZnO but also corroborates with the fact that HSE performed inadequately for transition and post-transition metal oxides \cite{PRB88p041107,JCC38p781}.

The challenge is, however, how to perform such a two-parameter calculation efficiently. This can be done by using the PP approach where the mixing parameters for the core and valence electrons are independently adjusted to accommodate the large spatial inhomogeneity between the $d$ and $sp$ orbitals. Let us consider first the simple case in Fig. 1(f) where HF PPs \cite{JCP122p014112,PRB77p075112,PRB97p085130,tanPCCP} are combined with PBE for valence electrons. Two different Zn PPs with 12 and 18 valence electrons (denoted as 12- and 18-val.) are considered here, respectively. The results show noticeable differences: increasing the number of valence electrons, $E_g$ decreases from 2.1 to 1.7 eV, while $E_d$ increases from 0.8 eV below the lower bound of the $d$-band set by experiment to 0.5 eV above the upper bound. If PBE for valence electrons is replaced by HSE as in Fig. 1(g), $E_g$ will increase to 3.5 (12-val.) and 3.1 eV (18-val.), respectively. For the $d$ bands, on the other hand, the effect of the HSE in the 12-val. case is insignificant, while in the 18-val. case, they are pushed down into the experimental range.

From the above discussion, three trends emerge: (1) band structure depends on the choice of PP. In particular, $E_g$ and $E_d$ are both inversely proportional to the number of valence electrons. (2) Replacing PBE by a hybrid functional increases $E_g$, but decreases $E_d$ although the effect is noticeable only in the 18-val. case. (3) For HSE with $\alpha=0.25$, the errors in $E_g$ and $E_d$ of all-electron [Fig. 1(b)] are opposite to those of 12-val. case [Fig. 1(g)]. The last point is particularly important as it suggests that a simultaneous correction of $E_g$ and $E_d$ can be achieved if one develop a 12-val. hybrid functional PP \cite{PRB97p085130}, instead of the HF PP. This is indeed the case as illustrated in Fig. 1(h) where (0.75, 0.25) for ($\alpha_c$, $\alpha_v$) have been used, as explained below.

\begin{figure}[tbp]
\includegraphics[width=0.65\columnwidth]{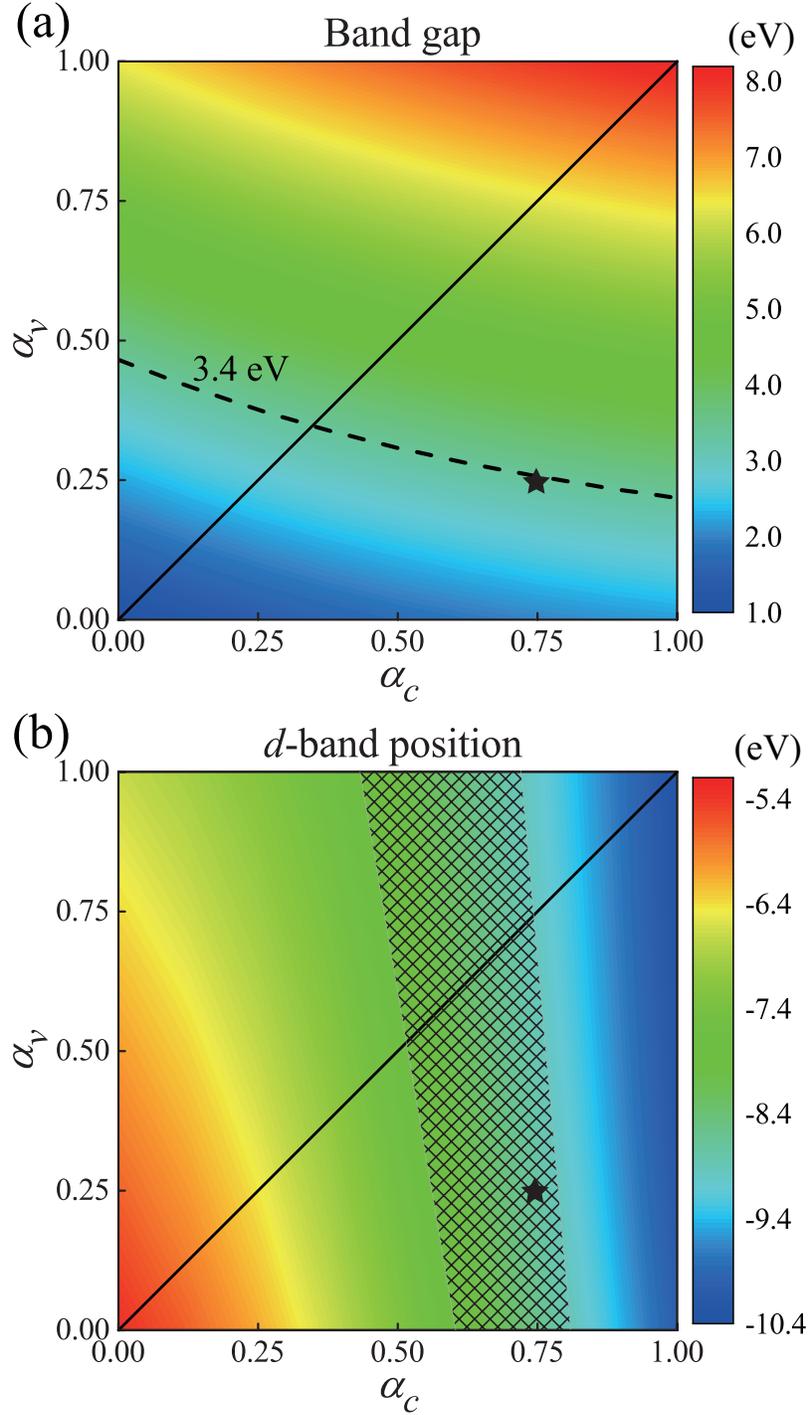}
\caption{Maps of (a) $E_g$ and (b) $E_d$ as functions of ($\alpha_c$, $\alpha_v$) for ZnO. Black dashed line in (a) represents the experimental $E_g$, whereas grid in (b) represents the experimental region for $E_d$. In (a) and (b), black diagonal lines are the allowed phase space if one is restricted to the all-electron single $\alpha$ scheme \cite{note1}, namely, $\alpha_c$ = $\alpha_v$ = $\alpha$, while black stars denote ($\alpha$$_c$, $\alpha$$_v$) = (0.75, 0.25), which fall within the experimental ranges of both $E_g$ and $E_d$.}
\end{figure}

Figure 2 shows for ZnO the $E_g$ and $E_d$ dependences on ($\alpha_c$, $\alpha_v$). It reveals that $E_g$ depends mainly on $\alpha_v$, while $E_d$ depends mainly on $\alpha_c$. To obtain the experimental $E_g$, $\alpha_v$ should be in the range of (0.23, 0.47) [See Fig. 2(a)]. To obtain the experimental $E_d$, $\alpha_c$ should be in the range of (0.43, 0.81). It is interesting to note that simultaneous corrections of $E_g$ and $E_d$ are obtained for $\alpha_v$ $\sim$ 0.25, which is deduced from perturbation theory \cite{JCP105p9982}. In this case the $\alpha_c$ is about 0.75. This pair of values, namely, ($\alpha_c$, $\alpha_v$) = (0.75, 0.25), is indicated in Fig. 2 by the black stars. Figure 1(h) shows the corresponding band structure.

Using the black diagonal lines $\alpha_v$ = $\alpha_c$ = $\alpha$ in Fig. 2, one can understand the inadequacy of the single $\alpha$ hybrid scheme more clearly \cite{note1}. The line intersects with experimental $E_g$ at $\alpha$ = 0.35. To intersect with the experimentally-determined $E_d$, however, $\alpha$ would have to be equal to or larger than 0.51.
This disparity effectively  characterizes the differences in the localization errors of the $sp$- and $d$-electrons. The difference of 0.16 (= 0.51 - 0.35) is significant, which implies that the inhomogeneity of the localization errors cannot be ignored.
For instance, when the correction for the $sp$-electrons is adequate, that for the $d$-electrons will be too small, leading to a too high $E_d$. By contrast, the HHF approach with two parameters $\alpha_c$ and $\alpha_v$ retains the adequate degrees of freedom to minimize the localization errors for both $sp$- and $d$-electrons. Although the $\alpha_c$ here does not directly affect the $sp$- or $d$-electrons in the outmost atomic shells, it affects the arrangement of the energy levels inside the core, subsequently the effective core$-$valence interactions. As such, the energies of the valence electrons are affected too. In such a real-space hybrid approach, $\alpha_c$ affects more the states closer to the nucleus, while $\alpha_v$ affects more the states that are spatially extended. They work together to produce adequate $E_g$ and $E_d$.

\begin{figure}[tbp]
\includegraphics[width=0.75\columnwidth]{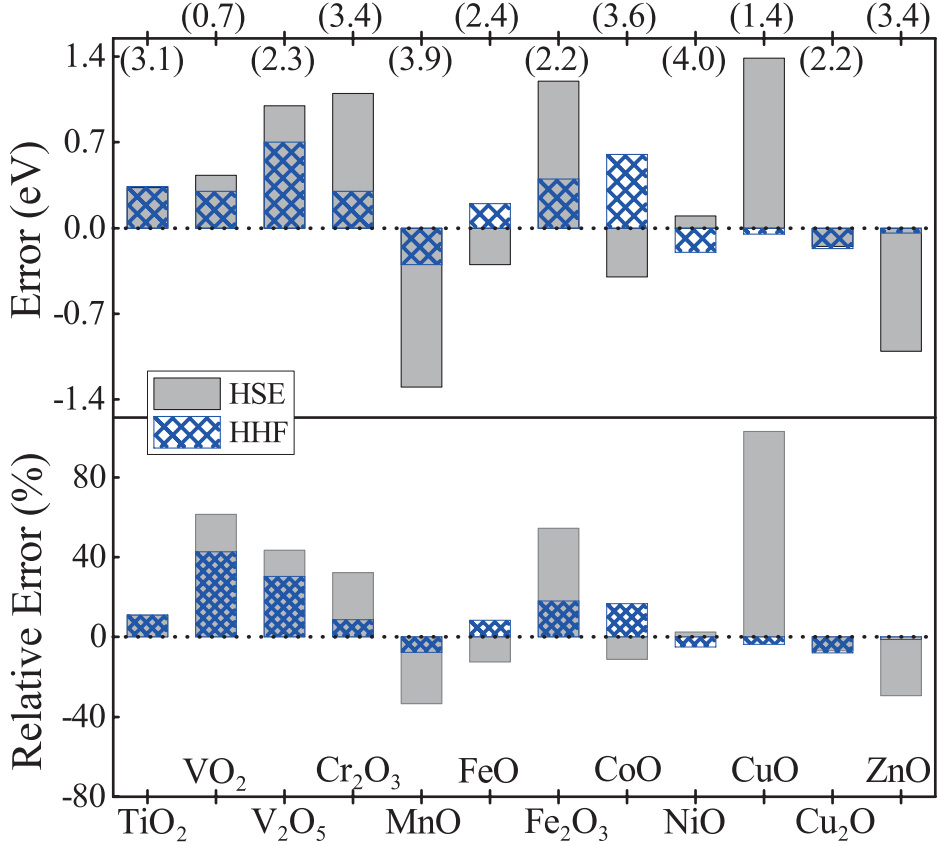}
\caption{(Upper panel) $E_g$ error and (lower panel) relative $E_g$ error of the HHF approach with ($\alpha$$_c$, $\alpha$$_v$) = (0.75, 0.25) and standard HSE with $\alpha$ = 0.25 for 12 $3d$ transition and post-transition metal binary oxides. Zero here means perfect agreement with experiment. Shown at the topmost of the figure in parentheses are the experimental band gaps (in eV) from Refs. \onlinecite{TiO2-Nanotech19p145605,VO2-PRB41p4993,V2O5-JPCS27p1237,Cr2O3-PRL55p418,MnO-FeO-CoO-NiO-PRB79p235114,MnO-FeO-CoO-NiO-PRB97p245132,FeO-JSSC12p355,Fe2O3-PRB79p035108,CoO-JPCS33p893,CuO-JAP53p1173,Cu2O-PSSB249p1487}. The HSE results are from Refs. \onlinecite{TiO2-JPCM24p195503,VO2-PRB85p115129,MnO-FeO-CoO-NiO-PRB79p235114,Fe2O3-JCP134p224706,CuO-Cu2O-PRB87p115111}. The experimental values for MnO and NiO display a large scattering. Here, the average values of 3.9 and 4.0 eV are used, respectively (See Table S2 of the Supplemental Material \cite{SM}).}
\end{figure}

Assuming that the localization errors are material-insensitive but orbital-sensitive as discussed here, the optimized ($\alpha_c$, $\alpha_v$) in ZnO, namely, (0.75, 0.25), should also apply to other $3d$ transition and post-transition metal oxides. As a test, we fix these parameters at the values of ZnO and calculate $E_g$ for 11 additional $3d$ transition and post-transition metal binary oxides. Figure 3 compares the errors of the current approach and standard HSE with $\alpha$ = 0.25 (For more details, see Table S2 of the Supplemental Material \cite{SM}), which are calculated using available experimental data as the reference.
The mean absolute error and relative error of HHF are 0.30 eV and 14.0\%, respectively, while those from HSE are 0.72 eV and 33.1\%. Clearly, the HHF approach shows a noticeable systematic improvement over HSE.

It is instructive to analyze the trends in Fig. 3. For example, a similar performance between HHF and HSE is obtained for Ti. Going to V and Cr, the errors of HHF decrease when compared to HSE. The same is true for Mn and Fe. For the late- and post-transition metals, the performance of HHF is even more remarkable. For CuO and ZnO (the two notorious cases of binary metal oxides), the absolute HHF errors are less than 0.1 eV, versus +1.4 and $-$1.0 eV of HSE, respectively. This chemical trend is expected to hold for other post-transition metal $d^0$ systems such as GaAs as its $d$ orbitals are also fully occupied. In general, HHF performs consistently better than HSE. The worst case is CoO \cite{noteNiO}. Even here, however, the HHF error of +0.6 eV is only slightly larger than the HSE error of $-$0.4 eV. Another observation is that hybrid functional PPs \cite{PRB97p085130} exhibit a remarkable transferability when multi-valency is involved, as evidenced in the results of V, Fe and Cu systems. This is not the case for HSE, as very different errors due to valency change appear unavoidably, e.g., $-$0.2 versus +1.4 eV for Cu$_2$O and CuO respectively.

\begin{figure}[tbp]
\includegraphics[width=0.75\columnwidth]{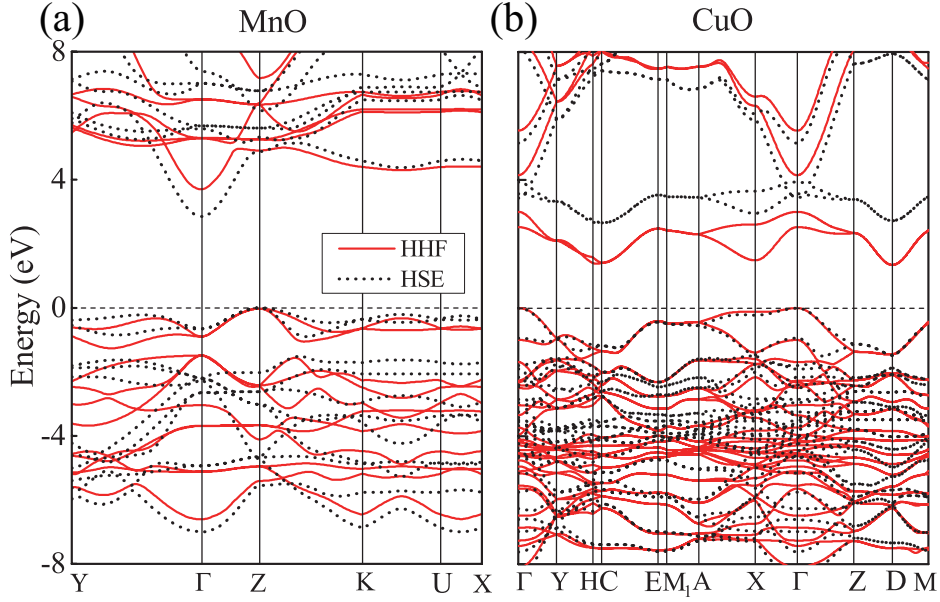}
\caption{Band structures of (a) MnO and (b) CuO. The HHF results with ($\alpha_c$, $\alpha_v$) = (0.75, 0.25) are given in red while those of standard HSE are given in dotted black. VBM is the energy zero. The labeling of the Brillouin zone follows the convention in Ref. \onlinecite{tanPCCP} for MnO and Ref. \onlinecite{CMS49p299} for CuO.}
\end{figure}

One can also look at the results in Fig. 3 from a different perspective. Because the standard HSE ignores the orbital difference, to fit experimental $E_g$ requires a system-dependent $\alpha$ for different $sp$-$d$ mixed compounds.
In contrast, the HHF approach eliminates such an $E_g$ dependence on $\alpha$ by capturing the error inhomogeneity with two mixing parameters. As it turns out, not only the $sp$ states, but also the 3$d$ states can be properly corrected. Interestingly, when the $d$ states are empty, the HHF result approaches that of HSE, as in the case of TiO$_2$ in Fig. 3. The same is expected for alkali-metal and alkali-earth-metal oxides.
For partially-occupied $d$ states, however, one has to take into account the following: (a) a strong oxygen $p$$-$metal $d$ orbital hybridization and (b) a crystal-symmetry-related splitting of the $d$ bands. As a result, $E_d$ is no longer as well-defined as in the case of $d^0$. Despite the complexity, the parameters ($\alpha_c$, $\alpha_v$) = (0.75, 0.25) produce nearly perfect $E_g$ for Cr$_2$O$_3$, MnO, Fe$_2$O$_3$ and CuO, without any additional adjusting parameters. Figure 4 selectively shows the band structures of MnO and CuO. It is interesting to note that here HHF and HSE produce very similar band dispersions, except for $E_g$. Noticeably, HHF increases $E_g$ for MnO but decreases $E_g$ for CuO. Thus, despite its simplicity, HHF should have captured the essential physics of transition and post transition-metal oxides.

In summary, the origin why the conventional hybrid functional calculations fail for transition and post-transition metal oxides is identified as its inability to characterize the orbital-dependent localization errors. We develop a PP-based HHF approach by introducing a discontinuity between the core and valence regions to compensate the different localization errors between the $sp$- and $d$-electrons at the same time. We show that the PP-based HHF approach improves the $E_g$ and $E_d$ of ZnO simultaneously and significantly. The same approach with the same mixing parameters also works for a whole range of transition and post transition metal oxides. This work thus offers a new prospect in terms of understanding the electron correlation phenomena such as magnetism and superconductivity in complex transition-metal oxides, as well as in band engineering for applications in electronics, photovoltaics, and catalysis.

\begin{acknowledgments}
H.T. thanks Dr. Jing Yang for helpful discussions. Work in China was supported by the Basic Science Center Project of NSFC (Grant Nos. 51788104), the Ministry of Science and Technology of China (Grant No. 2016YFA0301001), the National Natural Science Foundation of China (Grant Nos. 11674071, 11874089 and 11674188), the Beijing Advanced Innovation Center for Future Chip (ICFC), and the Beijing Institute of Technology Research Fund Program for Young Scholars. Work in the US (S.Z.) was supported by US DOE under Grant No. DE-SC0002623. S.Z. had been actively engaged in the design and development of the theory, participated in all the discussions and draft of the manuscript.
\end{acknowledgments}


\begin{thebibliography}{56}
\bibitem[{not({\natexlab{a}})}]{note1}
\bibinfo{note}{Strictly speaking, such diagonal lines are different from that
  of the ``consistent'' hybrid functional calculations, mainly originated from
  the approximations used for generating the hybrid functional PPs. It includes
  the neglect of relativistic effect, the difference between screened and
  unscreened hybrid functionals. Nevertheless, our test calculations show that
  such approximations yield an energy difference of about 0.2 eV for both $E_g$
  and $E_d$, hence not affecting the conclusion. Besides, the agreement between
  the previous ``consistent'' $\alpha$=0.375 [Fig. 1(c)] and our later
  $\alpha$$_c$=$\alpha$$_v$=0.35 [Fig. 2(a)] for the similar $E_g$, as well as
  the agreement between the ``consistent'' $\alpha$=0.50 [Fig. 1(d)] and our
  later $\alpha$$_c$=$\alpha$$_v$=0.51 [Fig. 2(b)] for the similar $E_d$, also
  prove the efficiency of such lines.}
\end{thebibliography}

\end{document}